\documentclass{article}
\usepackage{amsmath}
\usepackage{graphicx}

\begin{document}

\title{Strongly localized plasmon oscillations in a cluster of two
metallic nanospheres and their influence on spontaneous emission
of an atom}

\author{V.V. Klimov and D.V. Guzatov \\
P.N. Lebedev Physical Institute, Russian Academy of Sciences,\\
 53 Leninsky Prospekt, 119991 Moscow, Russia}

\date{\today}

\maketitle

\begin{abstract}

The plasmon oscillations in a cluster of two metallic nanospheres
are studied theoretically. Particular attention is paid to the
case of nearly touching spheres. Simple analytical expressions
have been found for the spectra of plasmon oscillations of
different symmetry in this case. A new type of the plasmon
oscillations, which are strongly localized between the spheres,
and which totally disappear at separation of the spheres, has been
discovered. The found plasmon oscillations have a dramatic effect
on optical properties of an atom localized between the spheres.

\end{abstract}

Much attention has been paid, of late, to the experimental and
theoretical study of optical properties of the metallic
nanoparticles. This interest is mainly due to a considerable
enhancement of local fields near the nanoparticles. An especially
high increase occurs in the case of plasmon polariton resonances
("plasmons") \cite{ref1,ref2} or  phonon polariton resonances
("phonons")\cite{ref36}. On the basis of this effect one considers
quite a number of possible applications. One of the most developed
is the use of large local fields for enhancement of the Raman
scattering cross-section \cite{ref3}. Recent experiments have
shown that such an increase may achieve 10-14 orders of magnitude,
which may help to resolve separate molecules
\cite{ref4}-\cite{ref6}. The local enhancement of the fields can
also be used to increase the fluorescence intensity and to
determine the structure of a single DNA strand without using the
fluorescent labels \cite{ref7,ref8}. By using the nanoparticles of
complex configuration one can provide enhancement of both the
absorption and the emission of light by natural and artificial
fluorophores \cite{ref9}. Of particular interest and promise are
the studies of optical properties of the clusters of two and more
metallic nanoparticles, because by changing the cluster's geometry
one can effectively control the spectra of the plasmon
oscillations. This effect makes it possible to produce, for
example, new types of biosensors \cite{ref10}-\cite{ref12}. A
whole series of experimental \cite{ref13}-\cite{ref16} and
theoretical \cite{ref17}-\cite{ref26} studies have been devoted to
the two-particle clusters.

Among the theoretical works one should note the paper
\cite{ref17}, where the method for defining the plasmon
oscillations of a two-sphere cluster on the basis of a plasmon
hybridization of separate spheres was proposed.

Most interesting is the case of nearly touching particles where
the enhancement of local fields is particularly high. This is the
most complicated case as well. In \cite{ref13}-\cite{ref16} this
interesting case was not actually considered, and the paper
\cite{ref18} being especially devoted to that problem, turned out
to be misleading. Basing on the numerical consideration of one of
the particular cases, the conclusion was made in \cite{ref19}
about failure of the Mie theory for the closely set nanospheres.

In the present work we will investigate the most interesting and
the most complicated case of the plasmon (phonon) oscillations in
the cluster of two almost touching metallic nanospheres. We shall
derive analytical expressions for the plasmon spectra in this
region. It will be shown that in this system there exists a new
type of plasmon oscillations (plasmonic molecule), which is
strongly localized between the spheres, and which cannot be found
within the framework of the hybridization method \cite{ref17}. We
shall also show that new modes can be excited only by a localized
light source (an excited atom, molecule, or a quantum dot) that
should lead to a dramatic changing of optical properties of an
atom. A geometry of the problem is illustrated in Fig.1. In the
case of $\varepsilon _{1} < 0$, $\varepsilon _{2} = 1$ the
geometry corresponds to two metallic nanoparticles, and in the
case of $\varepsilon _{1} = 1$, $\varepsilon _{2} < 0$ to two
spherical cavities (bubbles) in a metal. Hereinafter we  assume
that the dimensions of the studied particles are sufficiently
small in order one could use the longwavelength perturbation
theory \cite{ref34} but at the same time, they are essentially
large for one could neglect the non-local effects (space
dispersion of permittivity) \cite{ref33}. For simplicity here we
will consider only the case of equal nanospheres.

To determine the spectrum of the plasmon oscillations in the first
order of longwavelength perturbation theory \cite{ref34} one must
solve the equations:

\begin{equation}
\text{div}\mathbf{D}=0,\quad \text{rot}\mathbf{E}=0  \label{eq1}
\end{equation}

\noindent with usual boundary conditions of continuity of the
tangential component of $\mathbf{E}$ and the normal component of
 $\mathbf{D}$. To find further small corrections to spectrum of plasmon
oscillations one must use procedure elaborated in \cite{ref34}.
 By means of a conventional substitution, $\mathbf{E}
= - \nabla \Phi $, the system of equations (\ref{eq1}) is reduced
to the Laplace equation for the potential $\Phi $, which can be
conveniently solved in a bispherical system of coordinates. The
bispherical coordinates ($- \infty < \eta< \infty $, $0 < \xi \le
\pi $, $0 \le \varphi < 2\pi$) are connected with the Cartesian
coordinates by the relations \cite{ref27}

\begin{eqnarray}
x &=&a\frac{\sin \xi \cos \varphi }{\cosh \eta -\cos \xi },
\nonumber \\ y &=&a\frac{\sin \xi \sin \varphi }{\cosh \eta -\cos
\xi },  \nonumber \\ z &=&a\frac{\sinh \eta }{\cosh \eta -\cos \xi
}  \label{eq2}
\end{eqnarray}

 The surface $\eta=-\eta_{0}<0$ is the sphere of the radius
$R_{0}=a/\sinh\eta_{0}$. The other sphere can be set in the
similar way by the equality $\eta=\eta_{0}>0$. The dimensional
constant $a$ is the half distance between the poles of a
bispherical system of coordinates, and is determined by the next
expression: $a=\sqrt{R_{12}^{2}/4-R_{0}^{2}}$, where $R_{12}$ is
the distance between the centers of the first and second spheres.

The potential in space between the spheres ($-\eta_{0} < \eta <
\eta _{0} $) can be written in the form

\begin{align}
\Phi & =\frac{1}{a}\sqrt{\cosh \eta -\cos \xi
}\sum\limits_{n=0}^{\infty }\sum\limits_{m=0}^{n}P_{n}^{m}\left(
\cos \xi \right)  \nonumber \\ & \left\{ \left( \alpha _{mn}\cos
\left( m\varphi \right) +\beta _{mn}\sin \left( m\varphi \right)
\right) \cosh \left( \left( n+\frac{1}{2}\right) \eta \right)
\right.  \nonumber \\ & \left. +\left( \gamma _{mn}\cos \left(
m\varphi \right) +\delta _{mn}\sin \left( m\varphi \right) \right)
\sinh \left( \left( n+\frac{1}{2}\right) \eta \right) \right\}
\label{eq3}
\end{align}

\noindent where $\alpha _{mn} $, $\beta _{mn}$, $\gamma _{mn} $,
$\delta _{mn} $ are the coefficients to be determined.

For the potential inside the nanospheres we have, respectively,
($\eta < -\eta _{0} < 0$)

\begin{eqnarray}
\Phi ^{\left( 1\right) } &=&\frac{1}{a}\sqrt{\cosh \eta -\cos \xi }%
\sum\limits_{n=0}^{\infty }\sum\limits_{m=0}^{n}e^{\left( n+\frac{1}{2}%
\right) \eta }P_{n}^{m}\left( \cos \xi \right)   \nonumber \\
&&\left\{ a_{mn}^{\left( 1\right) }\cos \left( m\varphi \right)
+b_{mn}^{\left( 1\right) }\sin \left( m\varphi \right) \right\}
\label{eq4}
\end{eqnarray}

\noindent and ($\eta > \eta _{0} > 0$)

\begin{eqnarray}
\Phi ^{\left( 2\right) } &=&\frac{1}{a}\sqrt{\cosh \eta -\cos \xi }%
\sum\limits_{n=0}^{\infty }\sum\limits_{m=0}^{n}e^{-\left( n+\frac{1}{2}%
\right) \eta }P_{n}^{m}\left( \cos \xi \right)   \nonumber \\
&&\left\{ a_{mn}^{\left( 2\right) }\cos \left( m\varphi \right)
+b_{mn}^{\left( 2\right) }\sin \left( m\varphi \right) \right\}
\label{eq5}
\end{eqnarray}

\noindent in (\ref{eq4}) and (\ref{eq5}) $a_{mn}^{(1)}$,
$b_{mn}^{(1)}$ and $a_{mn}^{(2)}$, $b_{mn}^{(2)}$ are the
coefficients to be determined.

Expressions for the potentials in the form of
(\ref{eq3})-(\ref{eq5}) have the most general form, and make it
possible to determine \textit{all} the modes irrespective of their
symmetry.

 By using usual boundary conditions at the surface of each of the
nanospheres

\begin{eqnarray}
\Phi \left( \eta =-\eta _{0}\right)  &=&\Phi ^{\left( 1\right)
}\left( \eta =-\eta _{0}\right) ,  \nonumber \\ \varepsilon
_{2}\frac{\partial }{\partial \eta }\Phi \left( \eta =-\eta
_{0}\right)  &=&\varepsilon _{1}\frac{\partial }{\partial \eta
}\Phi ^{\left( 1\right) }\left( \eta =-\eta _{0}\right)
\label{eq51}
\end{eqnarray}

\noindent and

\begin{eqnarray}
\Phi \left( \eta =\eta _{0}\right)  &=&\Phi ^{\left( 1\right)
}\left( \eta =\eta _{0}\right) ,  \nonumber \\ \varepsilon
_{2}\frac{\partial }{\partial \eta }\Phi \left( \eta =\eta
_{0}\right)  &=&\varepsilon _{1}\frac{\partial }{\partial \eta
}\Phi ^{\left( 2\right) }\left( \eta =\eta _{0}\right)
\label{eq52}
\end{eqnarray}

\noindent and the recurrent relations for the Legendre functions
one can derive an infinite system of equations for all the
coefficients: $\alpha _{mn} $, $\gamma _{mn} $, $\beta _{mn}$,
$\delta _{mn} $, $a_{mn}$, and $b_{mn} $. In the most interesting
case of the two identical nanospheres these equations can be
written down in the following form

\begin{eqnarray}
&&\tau \left\{ \left( n-m\right) \tilde{\alpha}_{m,n-1}+\left(
\sinh \eta _{0}-\left( 2n+1\right) \cosh \eta _{0}\right)
\tilde{\alpha}_{mn}\right. \nonumber \\ &&\left. +\left(
n+m+1\right) \tilde{\alpha}_{m,n+1}\right\}   \nonumber \\
&=&-\left( n-m\right) \tanh \left( \left( n-\frac{1}{2}\right)
\eta _{0}\right) \tilde{\alpha}_{m,n-1}  \nonumber \\ &&+\cosh
\eta _{0}\left( \tanh \eta _{0}+\left( 2n+1\right) \tanh \left(
\left( n+\frac{1}{2}\right) \eta _{0}\right) \right)
\tilde{\alpha}_{mn} \nonumber \\ &&-\left( n+m+1\right) \tanh
\left( \left( n+\frac{3}{2}\right) \eta _{0}\right)
\tilde{\alpha}_{m,n+1}  \label{eq6}
\end{eqnarray}

\noindent and

\begin{eqnarray}
&&\tau \left\{ \left( n-m\right) \tilde{\gamma}_{m,n-1}+\left(
\sinh \eta _{0}-\left( 2n+1\right) \cosh \eta _{0}\right)
\tilde{\gamma}_{mn}\right. \nonumber \\ &&\left. +\left(
n+m+1\right) \tilde{\gamma}_{m,n+1}\right\}   \nonumber \\
&=&-\left( n-m\right) \coth \left( \left( n-\frac{1}{2}\right)
\eta _{0}\right) \tilde{\gamma}_{m,n-1}  \nonumber \\ &&+\cosh
\eta _{0}\left( \tanh \eta _{0}+\left( 2n+1\right) \coth \left(
\left( n+\frac{1}{2}\right) \eta _{0}\right) \right)
\tilde{\gamma}_{mn} \nonumber \\ &&-\left( n+m+1\right) \coth
\left( \left( n+\frac{3}{2}\right) \eta _{0}\right)
\tilde{\gamma}_{m,n+1}  \label{eq7}
\end{eqnarray}

\noindent where $\tau = \varepsilon _{1} / \varepsilon _{2} $ and
$\tilde { \alpha}_{mn} = \alpha _{mn} \cosh \left( {\left( {n +
{\frac{{1}}{{2}}}} \right)\eta _{0}} \right)$, $\tilde {\gamma}
_{mn} = \gamma _{mn} \sinh \left( {\left( {n + {\frac{{1}}{{2}}}}
\right)\eta _{0}} \right)$. Note that the recurrent equations for
the second series of the coefficients $\beta _{mn} $ and $\delta
_{mn} $ have the form (\ref{eq6}) and (\ref{eq7}) (the
substitutions $\alpha _{mn} \to \beta _{mn} $ and $\gamma _{mn}
\to \delta _{mn} $ are to be made).

The coefficients $a_{mn}^{\left( {1} \right)} $ and
$a_{mn}^{\left( {2} \right)} $ for the fields inside the
nanospheres can be found from the relations

\begin{eqnarray}
a_{mn}^{\left( 1\right) }e^{-\left( n+\frac{1}{2}\right) \eta _{0}} &=&%
\tilde{\alpha}_{mn}-\tilde{\gamma}_{mn},  \nonumber \\
a_{mn}^{\left( 2\right) }e^{-\left( n+\frac{1}{2}\right) \eta _{0}} &=&%
\tilde{\alpha}_{mn}+\tilde{\gamma}_{mn}  \label{eq8}
\end{eqnarray}

\noindent The equations for the coefficients $b_{mn}^{\left( {1}
\right)} $ and $ b_{mn}^{\left( {2} \right)} $ can be derived from
(\ref{eq8}) by means of the following substitutions
$a_{mn}^{\left( {1} \right)} \to b_{mn}^{\left( {1} \right)} $,
$a_{mn}^{\left( {2} \right)} \to b_{mn}^{\left( {2} \right)} $,
and $\alpha _{mn} \to \beta _{mn} $, $\gamma _{mn} \to
\delta_{mn}$.

The infinite systems of homogeneous equations (\ref{eq6}) and
(\ref{eq7}) will have nonzero solution only for special (resonant)
values of $\tau=\varepsilon_{1}/\varepsilon_{2}$. To find these
resonant values one should solve the generalized eigenvalue
problem. Note that the coefficients $\alpha _{mn} $ and $\gamma
_{mn} $ (which correspond to solution of different parity in
$\eta$) enter into (\ref{eq6}) and (\ref{eq7}) independently.

Figure 2 illustrates the dependence of the resonant permittivity
on distance between spheres for $m=1$. For any other $m$ we have
similar spectra. One can see in Fig.2 that there are three types
of the modes: \textit{L}-, \textit{M}- and \textit{T}-types. As
seen in Fig.2, if the distance between the nanospheres increases
the resonant permittivity for the \textit{T}- and \textit{L}-modes
takes the known \cite{ref22,ref25} values $\tau = - {\frac{{n +
1}}{{n}}}$, which correspond to the plasmon oscillations of
different electric multipolarity excited in a single sphere: $n=1$
corresponds to the dipole oscillations, $n=2$ corresponds to the
quadrupole oscillations and so on. Such modes have been well
studied and are described by hybridization of modes of the
separate spheres \cite{ref17}. As for the \textit{M}-modes, it
seems that their appearance had not been observed earlier. These
modes have a strongly localized character, because they can exist
in the clusters with ${\frac{{R_{12}}} {{2R_{0}}} } \leq 1.2$
only. Such modes can be considered as \textit{bound} states of
plasmons of the different spheres (plasmonic molecule). In order
to excite such modes one needs a localized source of
electromagnetic field. In the papers \cite{ref25,ref26} studying
the surface modes of a cluster one did not observe the appearance
of such modes because they considered an excitation of a cluster
by a plane electromagnetic wave which can not excite new
\textit{M}-modes.

From the above Figure 2 it is seen that the behavior of the
resonant permittivity in the region of the nearly touching spheres
becomes very complicated. To describe that complex behavior we
consider (\ref{eq6}) at the limit of $\eta _{0} \to 0$
(${\frac{{R_{12}}} {{2R_{0}}} } \to 1$). In this case we have
found an asymptotic expression for the resonance parameter $\tau $
(relative permittivity).

For the \textit{M}-modes we have

\begin{equation}
\tau _{m,M}=\frac{\varepsilon _{1}}{\varepsilon _{2}}=-\left(
M+\delta _{m}\right) \text{arccosh} \frac{R_{12}}{2R_{0}}+\ldots
\label{eq9}
\end{equation}

\noindent where $m=0,\,1,\,2,\,\ldots$; $M=1,\,2,\,3,\,\ldots$;
$\delta_{0}=1/2$ and $\delta_{1}\approx 0.914$.

For the \textit{L}-modes one can obtain the following analytical
solution

\begin{equation}
\tau _{m,L}=\frac{\varepsilon _{1}}{\varepsilon _{2}}=-\left( m+L-\frac{1}{2}%
\right)^{-1} /\text{arccosh} \frac{R_{12}}{2R_{0}}+\ldots
\label{eq11}
\end{equation}

\noindent where $m=0,\,1,\,2,\,\ldots$; $L=1,\,2,\,3,\,\ldots $ Of
course expressions (\ref{eq9}) and  (\ref{eq11}) are in agrement
with Fig.2 in the region ${\frac{{R_{12}}} {{2R_{0}}} } \to 1$.

From these expressions it follows that the resonant relative
permittivity of the cluster $\tau $ tends to zero
(\textit{M}-modes) or to the infinity (\textit{L}-modes) if the
nanospheres are drawing together. Tending of the relative
permittivity to zero corresponds to two physically different
cases, namely, two metallic nanospheres ($\varepsilon _{1} = 0$,
$\varepsilon _{2} = 1$) or two cavities in a metal ($\varepsilon
_{1} = 1$, $\varepsilon _{2} \to - \infty $). The analogous
situation occurs when the relative permittivity tends to the
infinity.

The eigenvectors of the \textit{L}- and \textit{M}-modes
correspond to the resonant permittivity (\ref{eq11}) and
(\ref{eq9}) at the limit of the closely set spheres can be
calculated analytically as well,

\begin{equation}
\tilde{\alpha}_{0,M}=\frac{1}{M}\left(
\begin{array}
[c]{c}%
\left.
\begin{array}
[c]{c}%
-1\\ \vdots\\ -1
\end{array}
\right\}  M\\ M\\ 0\\ \vdots\\ 0
\end{array}
\right)  ,\quad\tilde{\gamma}_{m,L}=\left(
\begin{array}
[c]{c}%
\left.
\begin{array}
[c]{c}%
0\\ \vdots\\ 0
\end{array}
\right\}  L-1\\ 1\\ 0\\ \vdots\\ 0
\end{array}
\right)  \label{eq12}%
\end{equation}

Note that the analogous expression for $\tilde {\delta} _{m,L} $
coincides with expression (\ref{eq12}) for $\tilde {\gamma} _{m,L} $, and $\tilde {%
\beta }_{0,M} = 0$. One can substitute these vectors into
(\ref{eq3})-(\ref{eq5}) to find the expressions for the potential
eigenfunctions.

For the eigenfunctions of the \textit{M}-mode ($m=0$) in space
between the spheres we obtain

\begin{align}
\left.  \Phi_{0,M}\right\vert _{\eta_{0}\rightarrow0}  & \approx-\frac{1}%
{Ma}\sqrt{\cosh\eta-\cos\xi}\sum_{n=0}^{M-1}\cosh\left(  \left(  n+\frac{1}%
{2}\right)  \eta\right)  P_{n}\left(  \cos\xi\right) \nonumber\\
& +\frac{1}{a}\sqrt{\cosh\eta-\cos\xi}\cosh\left(  \left(  M+\frac{1}%
{2}\right)  \eta\right)  P_{M}\left(  \cos\xi\right)  \label{eq13}
\end{align}

Inside the first sphere we have, correspondingly,

\begin{align}
\left. \Phi^{(1)} _{0,M}\right\vert _{\eta _{0}\rightarrow 0}& \approx -\frac{1}{Ma%
}\sqrt{\cosh \eta -\cos \xi }  \nonumber \\ &
\sum_{n=0}^{M-1}\left( 1+\left( n+\frac{1}{2}\right) \eta
_{0}\right) e^{\left( n+\frac{1}{2}\right) \eta }P_{n}\left( \cos
\xi \right)   \nonumber
\\
& +\frac{1}{a}\sqrt{\cosh \eta -\cos \xi }\left( 1+\left( M+\frac{1}{2}%
\right) \eta _{0}\right) e^{\left( M+\frac{1}{2}\right) \eta
}P_{M}\left( \cos \xi \right)   \label{eq14}
\end{align}

\noindent Inside the second sphere we obtain the analogous
expression.

For the eigenfunctions of the \textit{L}-mode in space between the
spheres and inside the first sphere we have, correspondingly,

\begin{eqnarray}
\left. \Phi _{m,L}\right\vert _{\eta _{0}\rightarrow 0} &\approx &\frac{1}{a}%
\sqrt{2\left( \cosh \eta -\cos \xi \right) }\sinh \left( \left( L-\frac{1}{2}%
\right) \eta \right)   \nonumber \\
&&P_{L-1}^{m}\left( \cos \xi \right) \sin \left( m\varphi +\frac{\pi }{4}%
\right)   \label{eq15}
\end{eqnarray}

\noindent and

\begin{eqnarray}
\left. \Phi^{(1)} _{m,L}\right\vert _{\eta _{0}\rightarrow 0} &\approx &-\frac{%
\eta _{0}}{a}\left( L-\frac{1}{2}\right) \sqrt{2\left( \cosh \eta
-\cos \xi \right) }e^{\left( L-\frac{1}{2}\right) \eta }
\nonumber \\
&&P_{L-1}^{m}\left( \cos \xi \right) \sin \left( m\varphi +\frac{\pi }{4}%
\right)   \label{eq16}
\end{eqnarray}

\noindent Inside the second sphere we get the analogous
expression.

Knowing the eigenfunctions one can calculate a surface charge
$\sigma ^{\left( {i} \right)}$ that is stored on each of the
nanospheres (\textit{i}=1, 2)

\begin{equation}
\sigma^{\left(  i\right)  }=\frac{\tau-1}{4\pi a}\left(
\cosh\eta_{0}-\cos
\xi\right)  (-1)^{i}\left.  \frac{\partial\Phi^{\left(  i\right)  }%
}{\partial\eta}\right\vert _{\eta=\eta_{i}}
\label{eq17}%
\end{equation}

Using the asymptotic expressions for the potentials (\ref{eq13})-(\ref{eq16}%
) one can write down the expression for the charge for the
\textit{M}-mode ($m=0$) in the form

\begin{eqnarray}
\sigma _{0,M}^{\left( 1\right) } &=&\sigma _{0,M}^{\left( 2\right)
}\approx \frac{\left( 1-\cos \xi \right) ^{3/2}}{8\pi a^{2}}
\nonumber \\ &&\left( -\frac{1}{M}\sum_{n=0}^{M-1}\left(
2n+1\right) P_{n}\left( \cos \xi \right) +\left( 2M+1\right)
P_{M}\left( \cos \xi \right) \right) \label{eq18}
\end{eqnarray}

For the \textit{L}-mode we get, respectively,

\begin{equation}
\sigma_{m,L}^{\left(  1\right)  }=-\sigma_{m,L}^{\left(  2\right)  }%
\approx\frac{\left(  1-\cos\xi\right)  ^{3/2}}{4\sqrt{2}\pi a^{2}}%
\frac{\left(  2L-1\right)  ^{2}}{2\left(  m+L\right)
-1}P_{L-1}^{m}\left(
\cos\xi\right)  \sin\left(  m\varphi+\frac{\pi}{4}\right)  \label{eq19}%
\end{equation}

We did not succeed in finding the analytical solution for the
\textit{T}-mode. Moreover, it is difficult to find limit at
${\frac{{R_{12}}} {{2R_{0}}} } \to 1$ of these curves. To
investigate this limit let us consider the limiting case of the
touching spheres (${\frac{{R_{12}}} {{2R_{0}}} } = 1$) within the
coordinate system of the touching spheres \cite{ref28}. In this
case, the spectrum of the plasmon oscillations that are
symmetrical to the \textit{z}=0 plane is determined by the
integral equation

\begin{eqnarray}
&&K_{m}\left( \lambda ^{\prime }\right) \int\limits_{0}^{\lambda
^{\prime }}d\lambda A_{m\lambda }I_{m}\left( \lambda \right)
+I_{m}\left( \lambda ^{\prime }\right) \int\limits_{\lambda
^{\prime }}^{\infty }d\lambda A_{m\lambda }K_{m}\left( \lambda
\right)   \nonumber \\ &=&\frac{1}{\tau -1}\left( \tau +\tanh
\left( \lambda ^{\prime }\right) \right) A_{m\lambda ^{\prime }}
\label{eq20}
\end{eqnarray}

\noindent where $I_{m} $ and $K_{m}$ are the modified Bessel
function and the Macdonald function, and $A_{m\lambda} $ is the
Hankel transform of the potential eigenfunction.

The integral equation (\ref{eq20}) for the eigenvalues $\tau
_{m,T} $ can be solved numerically

\begin{eqnarray}
\tau _{0,1} &\approx &-1.696,\quad \tau _{0,2}\approx -1.355,\quad
\tau _{0,3}\approx -1.237,  \nonumber \\ \tau _{1,1} &\approx
&-1.799,\quad \tau _{1,2}\approx -1.386,\quad \tau _{1,3}\approx
-1.182  \label{eq21}
\end{eqnarray}

Thus, in the case of the touching nanospheres the intrinsic
resonant values of \textit{T}-modes lie within the interval
[-2,-1]. The numerical solution of (\ref{eq21}) agrees well with
the functions represented in Fig.2 in the region of
${\frac{{R_{12}}} {{2R_{0}}} } \to 1$ (\textit{T}-modes).

In the case of antisymmetrical (longitudinal) \textit{L}-modes the
integral equation has the form:

\begin{eqnarray}
&&K_{m}\left( \lambda ^{\prime }\right) \int\limits_{0}^{\lambda
^{\prime }}d\lambda A_{m\lambda }I_{m}\left( \lambda \right)
+I_{m}\left( \lambda ^{\prime }\right) \int\limits_{\lambda
^{\prime }}^{\infty }d\lambda A_{m\lambda }K_{m}\left( \lambda
\right)   \nonumber \\ &=&\frac{1}{\tau -1}\left( \tau +\coth
\left( \lambda ^{\prime }\right) \right) A_{m\lambda ^{\prime }}
\label{eq23}
\end{eqnarray}

\noindent which seems to have only trivial solutions at finite
$\varepsilon_{1}/\varepsilon_{2} $. This result agrees with the
asymptotic of (\ref{eq11}) and disagrees with a numerical solution
found in \cite{ref19}. Perhaps, its difference from \cite{ref19}
is due to insufficient accuracy of the calculations in
\cite{ref19}.

Figure 3 illustrates surface charge distributions derived from an
accurate numerical solution of the recurrent equations (\ref{eq6})
and (\ref{eq7}) for $m=1$. For any other $m$ we have the similar
distributions. To plot these distributions we have used a polar
system of coordinates ($\rho$, $\theta$) for each sphere in plane
$y=0$. The surface charge distribution was depicted with $\rho
=R_{0}\left( 1+\tilde{\sigma }\left( \theta \right) \right) $
polar curve where $\tilde{\sigma }\left( \theta \right)$ is the
dimensionless surface charge density as a function of polar angle
of corresponding sphere. The \textit{M}-mode and its vividly local
character are of most interest. This mode is actually localized in
the region with characteristic size of the order of size of the
gap between the nanospheres. The \textit{T}- and \textit{L}-modes
are due to the charge oscillations over the whole surface of
spheres and are analogous to the charge distribution on separate
spheres. Note that the \textit{L}-modes have also appreciable
localization between the spheres.

It is very important that new \textit{M}- modes can not be excited
by uniform electric field. To excite such modes one must use
nonuniform field such as dipole radiation of an atom or molecule
placed near the cluster. On the other hand, the excited plasmon
modes will influence on decay rate and fluorescence of an atom or
molecule. The modification of decay rate by new modes can be used
to detect them.

The radiative decay rate $\gamma^{radiative}$ of an atom placed
near any nanobody can be described by \cite{ref29,ref30}

\begin{equation}
\frac{\gamma^{radiative}}{\gamma_{0}}=\frac{\left\vert \mathbf{d}%
_{total}\right\vert ^{2}}{d_{0}^{2}}\label{eq24}%
\end{equation}

\noindent where $\gamma _{0} $ is the atom spontaneous decay rate
in free space; $\mathbf{d}_{0} $, $\mathbf{d}_{total} =
\mathbf{d}_{0} + \delta \mathbf{d}$ , the dipole transition moment
and the total dipole moment of an atom + cluster, respectively.
In our case the expression for the dipole moment induced in the
cluster takes the form \cite{ref31}

\begin{eqnarray}
\delta d_{x} &=&-\sqrt{2}a\sum\limits_{n=1}^{\infty }n\left(
n+1\right) \left( \mathbf{d}_{0}\mathbf{\nabla }^{\prime }\right)
\alpha _{1n}, \nonumber \\ \delta d_{y}
&=&-\sqrt{2}a\sum\limits_{n=1}^{\infty }n\left( n+1\right) \left(
\mathbf{d}_{0}\mathbf{\nabla }^{\prime }\right) \beta _{1n},
\nonumber \\ \delta d_{z} &=&\sqrt{2}a\sum\limits_{n=0}^{\infty
}\left( 2n+1\right) \left( \mathbf{d}_{0}\mathbf{\nabla }^{\prime
}\right) \gamma _{0n} \label{eq25}
\end{eqnarray}

\noindent where $\mathbf{\nabla}^{\prime }$ means gradient over
the atom's coordinates; $\alpha _{1n} $, $\beta _{1n} $, $\gamma
_{0n} $ are the coefficients in (\ref{eq3}) found with taking into
account the atomic dipole source.

Figure 4 illustrates the dependence of the spontaneous decay rate
of an atom near two nanospheres (\ref{eq24}), (\ref{eq25}) on a
transition wavelength. The spheres are placed in vacuum
($\varepsilon_{2}=1$) and made of SiC, where phonon resonances can
be excited in the infrared \cite{ref35}. Strictly speaking SiC is
uniaxial hexagonal crystal with anisotropic tensor of dielectric
permittivity. However, strong TO$\bot$ (797 $\text{cm}^{-1}$) and
TO$\Vert$ (788 $\text{cm}^{-1}$) modes have very similar
dielectric properties. It allows us to consider SiC crystal in the
first approximation as isotropic one in the region of 788-797
$\text{cm}^{-1}$. In Fig.4 the influence of these modes on
spontaneous emission of an atom is shown. As seen in Fig.4(a), the
\textit{T}- and \textit{M}-modes can be excited by the atom with
the transversal dipole (the atom lies precisely between the
spheres). The excitation of only these modes can be explained by
the symmetry of an induced charge at the surface of the spheres,
in respect to the plane with $z = 0$. As the real part of the
permittivity increases (decrease in the transition wavelength)
then \textit{T}-modes are replaced by \textit{M}-modes, in
accordance with the functions of Fig.2, at transition over the
$\varepsilon_1=-1$ point. It is also important to mention that the
spontaneous decay rate for the \textit{M}-modes is substantially
higher than that for the \textit{T}-modes. In Fig.4(b) one can
observe only the \textit{L}-modes in the spontaneous decay rate
spectrum, provided that the dipole moment of an atom is oriented
along $z$-axis (the atom lies precisely between the spheres). This
fact is due to the symmetry of the excitation source also, i.e.
the charge distribution on spheres is to be antisymmetrical to the
plane with $z=0$. In the same Figure one can clearly see an
increase in the number of the resonances as real part of the
permittivity approaches $\varepsilon_1=-1$, which agrees with the
functions represented in Fig.2. As the transition wavelength
decreases (going over $\varepsilon_1=-1)$ no resonant phenomena
are observed for such an orientation of the dipole, except for the
dip at $\varepsilon_1=0$.

 Thus, in this paper the spectrum of plasmon oscillations occurring in a cluster of two
closely set nanospheres, or spherical bubbles, has been considered
in detail, and the respective analytical expressions have been
found. A new type of strongly localized plasmons oscillations has
been discovered. Such modes can be considered as \textit{bound}
states of plasmons of the different spheres (plasmonic molecule).
New modes exert influence on the optical properties of an atom
near a gap between spheres, and can be used to control the decay
rates of an atom and molecule. From the other hand well pronounced
dependencies of an atom optical properties on distance between
spheres, dipole moment orientation and permittivity can be used
for elaboration of new types of optical nanosensors.

\bigskip \textbf{Acknowledgement.} The authors are grateful to the Russian
Foundation for Basic Research (grant 04-02-16211) and the RAS
Presidium program ''The influence of atomic-crystalline and
electron structure on the properties of condensed media'' for the
partial financial support of the present work.

\pagebreak

\newpage
\begin{center}
\bigskip {\LARGE List of Figure Captions}
\end{center}

Fig.1 Geometry of the problem.

Fig.2 Resonant permittivity as the function of normalized distance
between two identical spheres ($m=1$).

Fig.3 Surface charge density (in the relative units) for different
modes with $m=1$ (the plane $y=0$ is shown). (a) \textit{L}-modes;
(b) \textit{T}-modes; (c) \textit{M}-modes. The distance between
spheres is ${\frac{{R_{12}}} {{2R_{0}}} } = 1.1$. The broken line
denotes a surface of the sphere; the colored lines correspond to
the mode indices: red-1; green-2; blue-3 (cf. Fig.2).

Fig.4 The spontaneous emission radiative decay rate of an atom
near the two nanospheres made of SiC. The spheres have the radii
of 50 nm; the distance between their centers is of 101 nm. The
atom is placed precisely between the spheres and its dipole moment
is directed: (a) transversally to the rotation axis; (b)
longitudinally to the rotation axis.

\pagebreak
\newpage

\begin{figure}
\centering
\includegraphics[width=8cm]{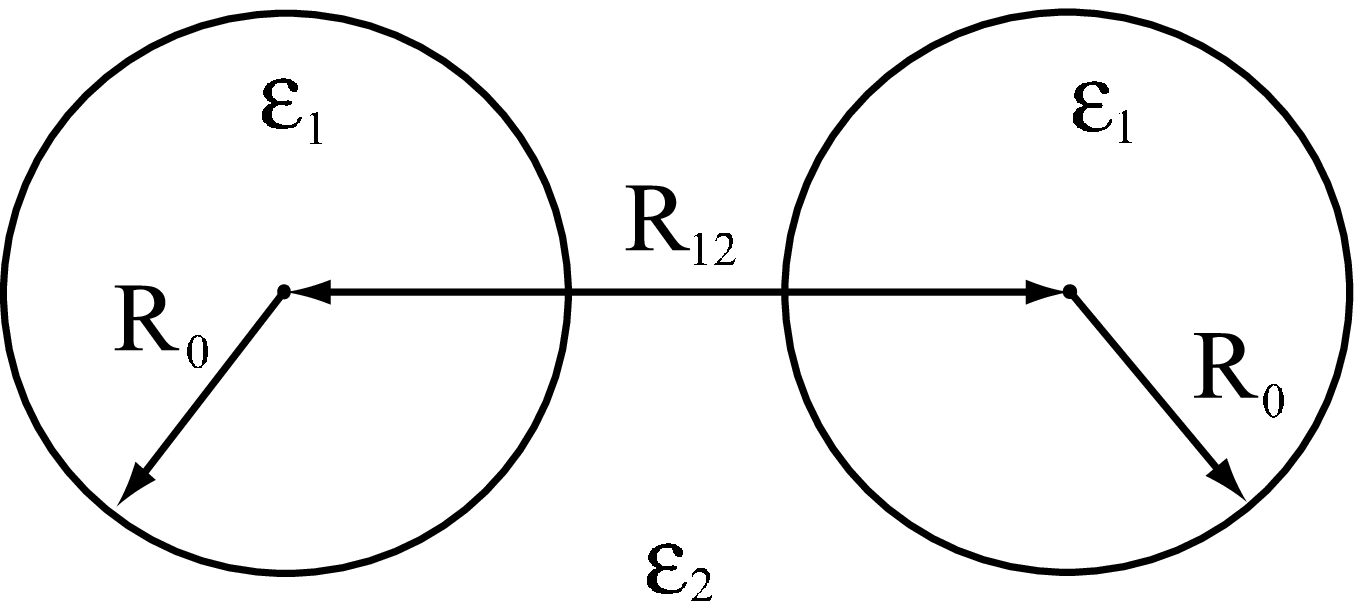}
\caption{}
\end{figure}

\pagebreak
\newpage

\begin{figure}
\centering
\includegraphics[width=10cm]{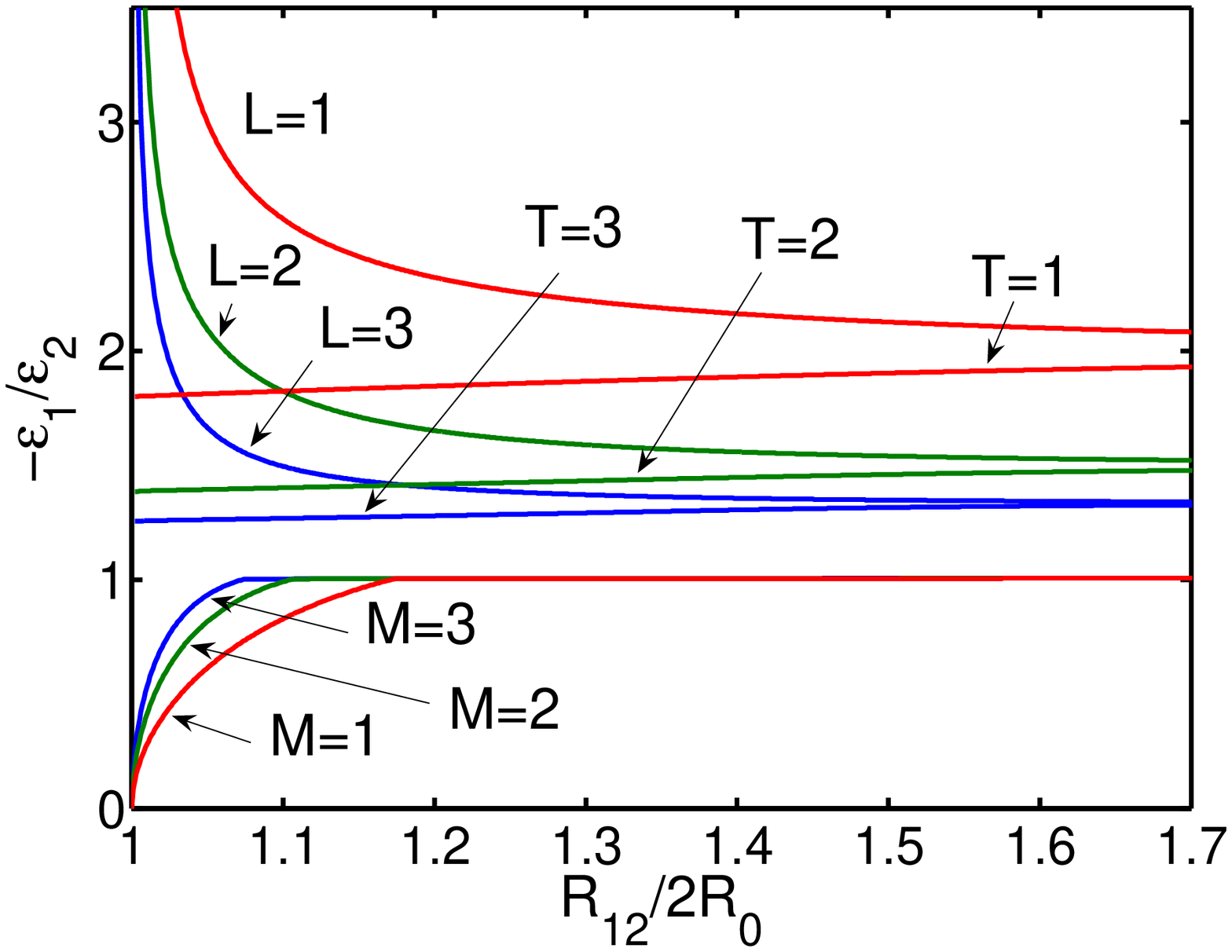}
\caption{}
\end{figure}

\pagebreak
\newpage

\begin{figure}
\centering
\includegraphics[width=9cm]{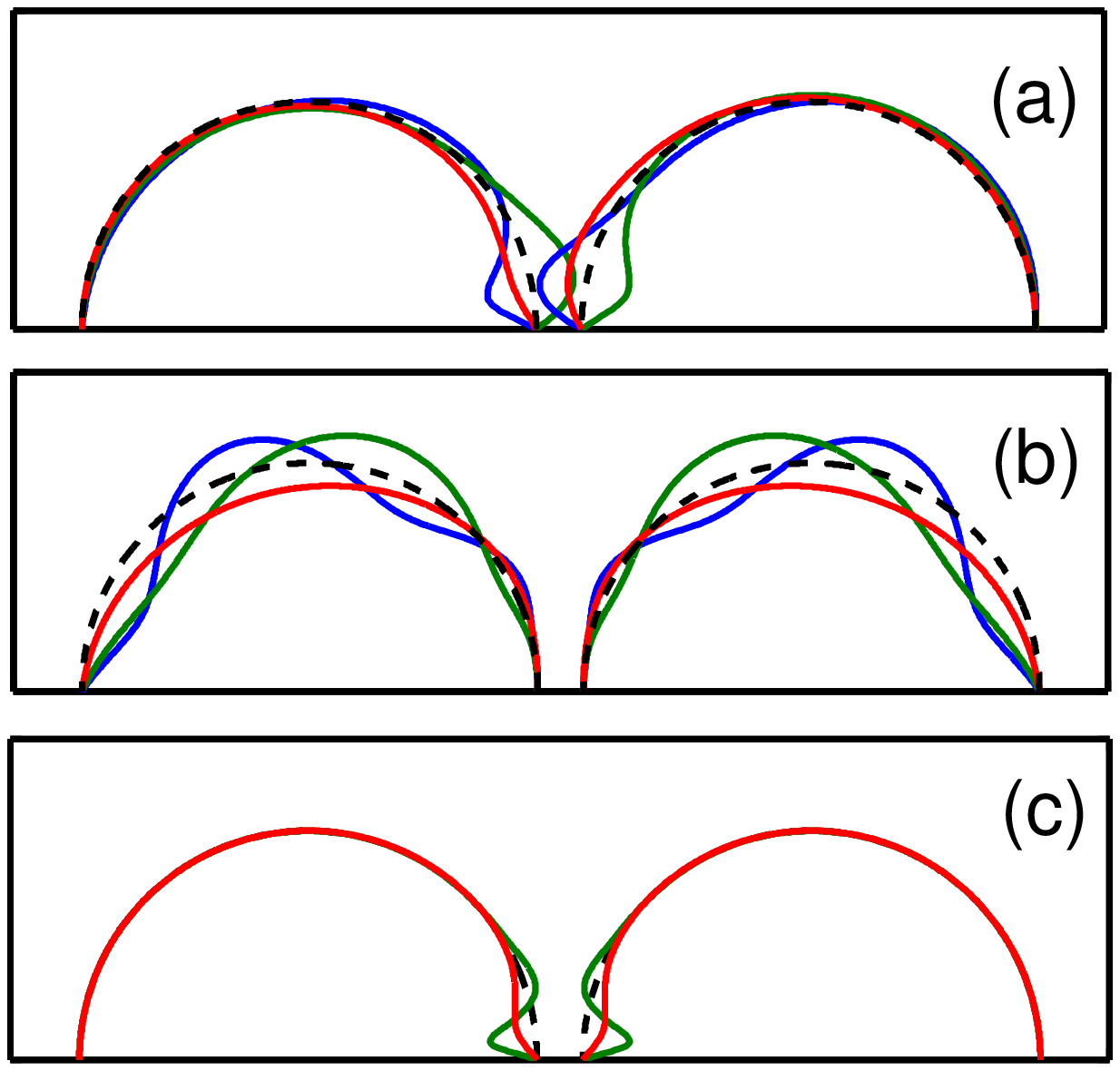}
\caption{}
\end{figure}

\pagebreak
\newpage

\begin{figure}
\centering
\includegraphics[width=10cm]{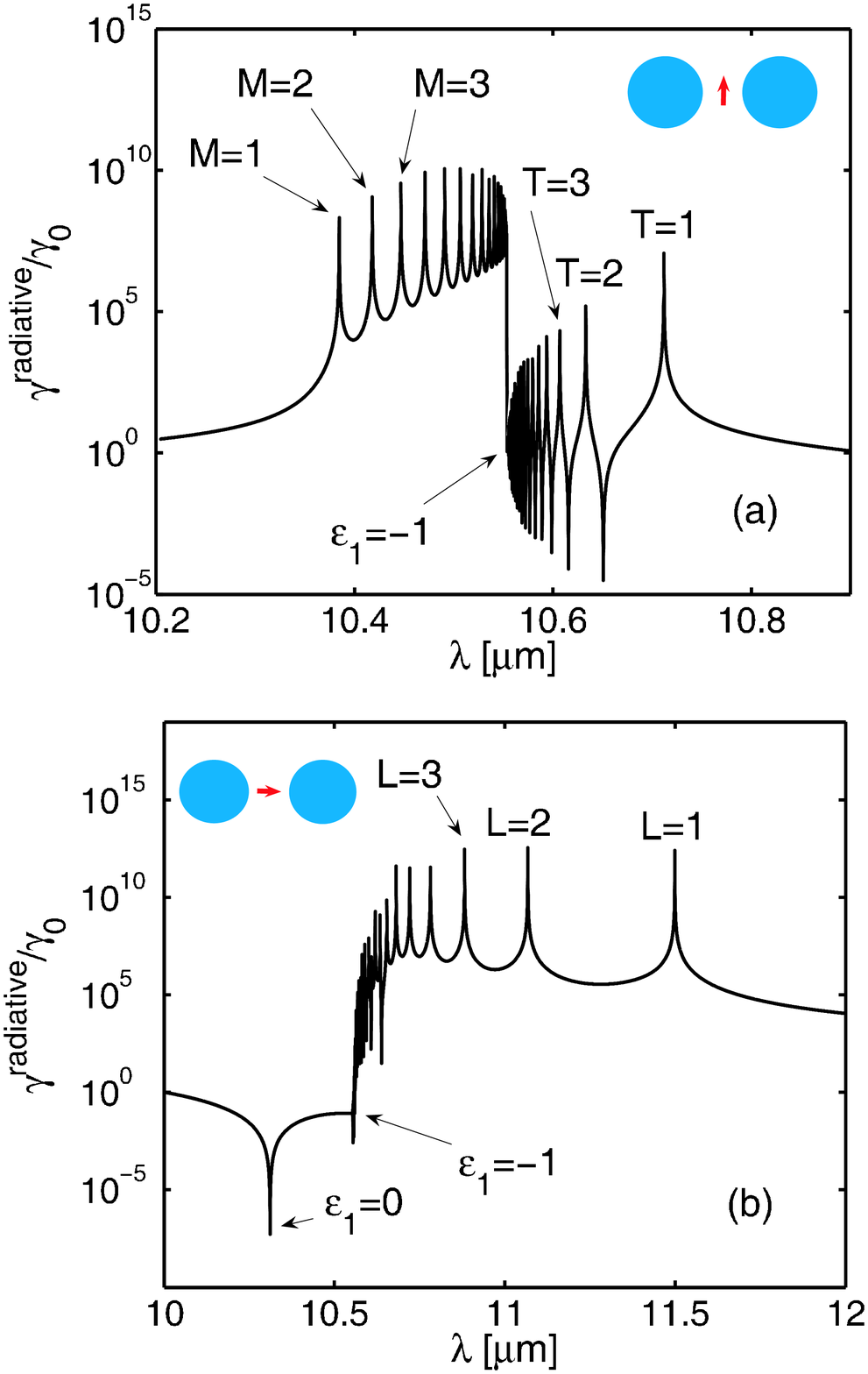}
\caption{}
\end{figure}

\end{document}